\documentclass[final,3p,10pt]{elsarticle}

\usepackage{lineno, amsmath, amssymb}
\usepackage{hyperref}
\usepackage[ruled,linesnumbered,lined]{algorithm2e}
\usepackage{tikz}
\usetikzlibrary{arrows,automata}
\usepackage{subfig}
\newcommand{\npat}{r} 
\newcommand{\data}{data} 
\newcommand{\XOR}{CNOT} 
\newcommand{\NOT}{X} 
\newcommand{\ip}{s} 
\usepackage{color}

\input{Qcircuit}

\modulolinenumbers[5]

\journal{Neurocomputing}

\begin{document}

\begin{frontmatter}

\title{Parametric Probabilistic Quantum Memory}

\author[dc-ufrpe]{Rodrigo~S.~Sousa}
\author[dc-ufrpe]{Priscila~G.~M.~dos~Santos}
\author[cin-ufpe]{Tiago~M.L.~Veras}
\author[deinfo-ufrpe]{Wilson~R.~de~Oliveira}
\author[cin-ufpe]{Adenilton~J.~da~Silva}
\cortext[mycorrespondingauthor]{A. J. da Silva}
\ead{ajsilva@cin.ufpe.br}

\address[cin-ufpe]{Centro de Inform\'{a}tica, Universidade Federal de Pernambuco}	
\address[dc-ufrpe]{Departamento de Computa\c{c}\~{a}o, Universidade Federal Rural de Pernambuco}
\address[deinfo-ufrpe]{Departamento de Estat\'{i}stica e Inform\'{a}tica, Universidade Federal Rural de Pernambuco}

\begin{abstract}
Probabilistic Quantum Memory (PQM) is a data structure that  computes the 
distance from a binary input to all binary patterns stored in superposition on 
the memory. This data structure allows the development of 
heuristics to speed up artificial neural networks architecture selection. In this work, 
we propose an improved parametric version of the PQM to perform 
pattern classification, and we also present a PQM quantum circuit suitable for Noisy
Intermediate Scale Quantum (NISQ) computers. We present a classical evaluation of a 
parametric PQM network classifier on public benchmark datasets. We also perform experiments 
to verify the viability of PQM on a 5-qubit quantum computer. 
\end{abstract}


\end{frontmatter}

\section{Introduction}

Quantum Computing is a computational paradigm that has been harvesting increasing attention for decades now. 
 Several quantum algorithms have time advantages over their best known classical counterparts~\cite{shor1999polynomial,grover1997quantum, rebentrost2014quantum,dosSantos2018quantum}.
The current advances in quantum hardware are bringing us to 
the era of Noisy Intermediate-Scale Quantum  (NISQ) computers~\cite{preskill2018quantum}.
The quest for quantum supremacy is the search for an efficient solution of a task in a quantum computer that current classical computers are not able to efficiently solve.
Some authors argue that given the current state of the art, we will achieve quantum supremacy in the next few years~\cite{harrow2017quantum}.
One of the approaches to achieve this supremacy and to expand the potential applications of quantum computers is through quantum machine learning~\cite{biamonte2017quantum}.

Machine learning (ML)~\cite{mitchell} aims at developing automated ways for computers to learn a specific task from a given set of data samples. 
ML has several applications in image classification~\cite{SHI20191}, NP-hard problems~\cite{ZHANG2019166} and others. Some of the limitations of ML are the lack of algorithms to select an ML method and to avoid local minimums.

Quantum machine learning investigates the use of quantum computing concepts to build quantum-enhanced machine learning models able to outperform the classical ones~\cite{biamonte2017quantum}. 
For instance, quantum computing can be used to efficiently convert a nonlinearly separable dataset into a linearly separable one by exploiting the exponential size of quantum spaces~\cite{havlivcek2019supervised} 
and to select neural networks architectures without dependence of weights initialization~\cite{dasilva2016quantum, fawaz2019training}.
Quantum machine learning proposes  speedups of some algorithms used in machine learning such as reinforcement learning~\cite{dunjko2016quantum} and systems of linear equations~\cite{harrow2009quantum}. 

There are several works proposing quantum machine learning models such as a quantum generalization of a neural network~\cite{wan2017quantum},
a quantum distance-based classifier~\cite{schuld2017implementing},
and quantum associative memories~\cite{ventura2000quantum, ezhov2000quantum, trugenberger2001probabilistic, 
trugenberger2002quantum}. In~\cite{trugenberger2001probabilistic, 
trugenberger2002quantum} is introduced a Probabilistic Quantum Memory (PQM) that  computes the Hamming distance from an $n$-bits input pattern to $k$ $n$-bits  patterns with computational cost linear in the number of bits, $O(n)$.

However, the PQM~\cite{trugenberger2001probabilistic} (and quantum associative memories in general) has received criticism from various authors. 
The memory state is lost in every execution of the retrieval algorithm when the memory state is measured. 
Memory collapse is considered to be the main limitation of this memory~\cite{PhysRevLett.91.209801} as argued in~\cite{schuld2014quantum} since the $O(m)$ cost to reload $m$ patterns in the memory using its storage algorithm jeopardizes the PQM advantages. 
Another author~\cite{dunjko2018machine} claims that the probabilistic quantum memory cannot be considered as a complete model. 
However, in~\cite{dosSantos2018quantum} we show a PQM application in which a single execution of the retrieval algorithm is sufficient to evaluate, probabilistically, the artificial neural networks architectures without the need for weights initialization. We used the PQM as a data 
structure to store and train artificial neural networks in superposition and to devise a quantum algorithm able to evaluate and select neural network architectures.

The objective of this work is to improve the PQM model in pattern classification tasks extending the model described in~\cite{dosSantoswnn}. To accomplish this aim, we propose two approaches:
\begin{enumerate}
\item We introduce the Parametric PQM (P-PQM), where the parameter allows the PQM to compute a weighted Hamming distance and adapt the model to the given training dataset.
 \item We also propose a hybrid classical-quantum version of the PQM retrieval algorithm suitable for NISQ computers.
\end{enumerate}

The remainder of this work is structured as follows. Section~\ref{sec:qc} gives the concepts of quantum computing to make this work self-contained. Section~\ref{sec:relatedworks} describes related works and the probabilistic quantum memory, and Section~\ref{sc:QWC} presents a quantum weightless classifier and one of its limitations. Section~\ref{sec:pqwc} and Section~\ref{sec:qubit-redc} show the main contributions of this work. In Section~\ref{sec:pqwc}, we describe the proposed parametric probabilistic quantum memory and present simulated experiments to evaluate a weightless neural network based on the P-PQM. In Section~\ref{sec:qubit-redc}, we present a modification of the PQM that allows its implementation in a NISQ computer. Finally, in Section~\ref{sc:conclusion} we draw some conclusions and list some possible further works.

\section{Quantum Computing}\label{sec:qc}

Quantum computing is a field that is receiving increasing attention due to its current advances~\cite{preskill2018quantum}. 
It is a field that uses concepts and results from quantum mechanics and computing theory. 
A quantum computer is a computational device capable of representing and manipulating information at the quantum level to perform computational tasks~\cite{nielsen2002quantum}. 
In quantum computing, the quantum bit (qubit) represents the basic unit of information, representing a two-level quantum system. Analogously to the behavior of a subatomic particle, the qubit can be in more than one \emph{basis} state at a given time.
Eq.~\eqref{eq1} describes one qubit in superposition (\emph{linear combination}), where
$\ket{0}$ and $\ket{1}$ are orthonormal vectors (basis), $\alpha$ and $\beta$ are the probabilistic amplitudes (\emph{complex numbers}) associated with the states $\ket{0}$ and $\ket{1}$, and $|\alpha|^2 + |\beta|^2 = 1$.

\begin{equation}
\ket{\psi} = \alpha \ket{0} + \beta \ket{1} \label{eq1}
\end{equation}

As is customary in quantum computing, the basis is fixed to $\{\ket{0}, \ket{1}\}$, and is called the \emph{computational basis}, where 
\[\ket{0} = \begin{bmatrix} 1 \\ 0 \end{bmatrix} \mbox{and } \ket{1}= \begin{bmatrix} 0 \\ 1 \end{bmatrix}.\]
Quantum gates are used to modify the state of a quantum system. 

In the computational basis, a gate is represented by a unitary matrix and carries out a reversible operation on the quantum state.

The Pauli gates are examples of quantum operators. They are defined by the matrices below. 

 $$X = \begin{bmatrix}
0   &    1 \\
1    &   0 \\
\end{bmatrix}; 
Y = \begin{bmatrix}
0   &    i \\
-i    &  0 \\
\end{bmatrix};   Z = \begin{bmatrix}
1   &    0 \\
0    &  -1 \\
\end{bmatrix}$$

The Hadamard gate $H$ (Eq.~\eqref{eq:H}) can create a state 
superposition; where $H\ket{0} = 
\frac{1}{\sqrt{2}}(\ket{0} + \ket{1})$ and $H\ket{1} = 
\frac{1}{\sqrt{2}}(\ket{0} - \ket{1})$. 

\begin{equation}
\label{eq:H}
H = \frac{1}{\sqrt{2}}\begin{bmatrix}
    1   &    1 \\
    1    &  -1 \\
\end{bmatrix}
\end{equation}

Controlled gates operate on a \emph{target} qubit depending on the state of another qubit used as \emph{control}. 

The CNOT gate is the controlled version of the $X$ gate and flips target qubit if the control qubit is in the state $\ket{1}$. 
Eq.~\eqref{eq:cnot} describes the CNOT gate.

\begin{equation}
\label{eq:cnot}
CNOT = \begin{bmatrix}
1   &    0  &  0 & 0\\
0   &    1  &  0 & 0\\
0   &    0  &  0 & 1\\
0   &    0  &  1 & 0\\
\end{bmatrix}
\end{equation}

An important characteristic of quantum systems is the need to measure for extracting information from a quantum state. 
After a measurement, the system collapses to one of its superposed basis states. 
Given the quantum state described in Eq.
~\eqref{eqpsi}, the probability of finding $\ket{i}$ after a measurement is
$p_i = |\alpha_i|^2$ and the state will collapse to $\ket{i}$. 

\begin{equation}
\ket{\psi} = \sum_i \alpha_i \ket{i} \label{eqpsi}
\end{equation}

Due to the capacity of dealing with states in superposition and other incorporated quantum effects (such as \emph{entanglement}), 
quantum computers provide an alternative way of computing, which presumably could be used to solve problems 
in a way that presumably has no counterpart in classical computing~\cite{shor1999polynomial}.

\section{Related works}
\label{sec:relatedworks}
The first model of Quantum Associative Memory (QAM) was proposed in~\cite{ventura2000quantum} and used a variation of Grover's algorithm. 
The main idea of QAM is to store patterns in superposition, allowing to store $2^n$ patterns using $n$ qubits.
The use of Grover's algorithm is one limitation of the first QAM that searches for exact patterns and not similar patterns.
The QAM performs a search for exact patterns and not similar patterns, and the required number of quantum operators limit the use of QAM in NISQ computers.
A Probabilistic Quantum Memory (PQM) was proposed in~\cite{trugenberger2001probabilistic}. 
The retrieval algorithm of the PQM is not based on Grover's algorithm and searches for similar patterns instead of performing an exact search.
However, the retrieval algorithm is probabilistic and depends on the distribution of patterns stored in the memory.  

Quantum associative memories have been used to perform classification tasks in several works~\cite{zhou2010quantum, schuld2017implementing, schuld2014quantum, dosSantos2018quantum, de2019quantum, singh2017classification, njafa2018quantum}. 
In~\cite{singh2017classification} Grover's algorithm and a quantum associative memory based on it are used to perform classification tasks in a toy dataset representing orange and apples with 3-qubit patterns. 
In~\cite{schuld2014quantum} the PQM is used to classify digits from the MNIST dataset, the PQM accuracy on the test set is only 50\%. 
An evaluation of the PQM in benchmark classification datasets was performed in~\cite{dosSantoswnn}. One limitation of the PQM is the lack of parameters to adjust the model to a dataset~\cite{dosSantoswnn, schuld2014quantum}.
 
\subsection{Probabilistic Quantum Memories}\label{sc:PQMC}

In this section, we present the quantum memory model used to build a weightless network classifier.
The Probabilistic Quantum Memory (PQM)~\cite{trugenberger2001probabilistic, trugenberger2002quantum} is a content-addressable quantum memory.
It outputs the probability of a given input pattern being stored in the memory by
calculating the Hamming distance between the input pattern and all the patterns stored in the memory. It is a probabilistic model designed to recognize incomplete or noisy information. Despite being an associative
model, the PQM possess a highly scalable storage capability, being able to
store all the possible $2^n$ binary patterns of $n$ bits. 
The following subsections explain the PQM storage and retrieval algorithms.

\subsubsection{The Storage algorithm}

The storage algorithm receives a dataset $\data = \cup_{i=1}^\npat \{p^i\}$ with $\npat$ patterns each with $n$ bits,  uses three quantum registers and follows Algorithm~\ref{alg:storing} to produce state $\ket{M}$ described in Eq.~\eqref{eq5}.
\begin{equation}
\ket{M} = \frac{1}{\sqrt{\npat}} \sum_{i=1}^\npat \ket{p^i} \label{eq5}
\end{equation}
 Before being processed and stored on the 
memory, every pattern must be initialized in an input register $\ket{p}_n$. The 
memory itself is separated from the input patterns which have not yet been 
processed and resides in the register $\ket{m}_n$. After the end of the storage algorithm,
the register $\ket{m}_n$ will be in the state $\ket{M}$; 
$\ket{m}_n$ refers to the memory quantum register during the construction of 
the uniform superposition and $\ket{M}$ is the resulting memory 
state. The last quantum register used is an auxiliary two-qubit register 
$\ket{u}_2=\ket{u_1u_2}=\ket{u_1}\otimes\ket{u_2}$, $u_i\in\{0,1\}$, which is used to keep tabs on which patterns are already stored and 
which still need to be processed and written.  The algorithm initial state $\ket{\psi_0^1}$ using the three registers is shown 
in Eq.~\eqref{eq6}.

\begin{equation}
\ket{\psi_0^1} = \ket{p_1p_2 \cdots p_n;u_1u_2;m_1m_2 \cdots m_n} 
\label{eq6}
\end{equation}

\begin{algorithm}
    Prepare the initial state $\ket{\psi_0^i} = 
    \ket{0_1,\cdots,0_n;01;0_1,\cdots,0_n}$ \\ \label{str:initial}
    \ForEach{$p^i \in \data$}{ \label{str:loop}
        Load $p^i$ into quantum register $\ket{p}_n$ \\ \label{str:load}
        $\ket{\psi_1^i} = \prod_{j=1}^n 
        2\XOR_{p_j^i,u_2,m_j}\ket{\psi_0^i}$ \\ \label{str:step1}
        $\ket{\psi_2^i} = \prod_{j=1}^n \NOT_{m_j} 
        \XOR_{p_j^i,m_j}\ket{\psi_1^i}$ \\ \label{str:step2}
        $\ket{\psi_3^i} = n\XOR_{m_1\cdots m_n,u_1}\ket{\psi_2^i}$ \\ 
        \label{str:step3}
        $\ket{\psi_4^i} = CS_{u_1,u_2}^{\npat+1-i}\ket{\psi_3^i}$ \\ 
        \label{str:step4}
        $\ket{\psi_5^i} = n\XOR_{m_1\cdots m_n,u_1}\ket{\psi_4^i}$ \\ 
        \label{str:step5}
        $\ket{\psi_6^i} = \prod_{j=1}^n \XOR_{p_j^i,m_j} 
        \NOT_{m_j}\ket{\psi_5^i}$ \\
        $\ket{\psi_7^i} = \prod_{j=1}^n 
        2\XOR_{p_j^i,u_2,m_j}\ket{\psi_6^i}$ \\
        Unload $p^i$ from quantum register $\ket{p}_n$ \\
    }
    \caption{Probabilistic quantum memory storage algorithm}
    \label{alg:storing}
\end{algorithm}

Algorithm~\ref{alg:storing} describes the storage algorithm receiving as input a dataset
with  $\npat$ $n$-bit patterns. Step~\ref{str:initial} initializes the quantum registers $\ket{p}_n\ket{u}_2\ket{m}_n$ with the quantum state $\ket{0}_n\ket{01}\ket{0}_n$.  The 
second qubit in $\ket{u}_2$ indicates whether a pattern has been already stored 
or not. In this case, a $\ket{1}$ in the second qubit of $\ket{u}_2$, \emph{.i.e.} $\ket{u_2}=\ket{1}$, indicates that 
the pattern has not been stored yet.

The for loop in line~\ref{str:loop} is repeated for each pattern $p^i$ in $\data$. Step~\ref{str:initial} initializes the quantum register $\ket{p}_n$ with a pattern $p^i$ from $\data$. Step~\ref{str:step1} uses $n$ 
$2\XOR$ operations to make a copy of the $n$ bits from the pattern in $\ket{p}_n$ 
to the respective $\ket{m}_n$ register flagged by $\ket{u_2} = \ket{1}$. The $2\XOR$ 
operation is equivalent to a Toffoli gate  that flips the target bit if the two control bits are in state $\ket{1}$. 

Step~\ref{str:step2} applies  $\XOR$ operations to 
registers $\ket{p}_n$ and $\ket{m}_n$ followed by an $\NOT$ operation to 
$\ket{m}_n$. This step fills with $1$s all the bits in the memory register 
which are equal to the respective bits in register $\ket{p}_n$; this is true 
only for the pattern which is being currently processed.

Line~\ref{str:step3} uses the $n\XOR$ operation, which is a generalization of 
the $CNOT$ gate for $n$ bits. The operation is controlled by all the bits of 
$\ket{m}_n$ and is applied to the first bit of $\ket{u}_n$. Thus, if $\ket{m}_n 
= \ket{1}_n$ the first bit of $\ket{u}_2$ is flipped.

Step~\ref{str:step4} adds the input pattern $p^i$ to the memory register 
with uniform amplitudes. This is done by applying the $CS^j$ gate, shown below:

$$CS^j = \begin{bmatrix}
1 & 0 & 0 & 0 \\
0 & 1 & 0 & 0 \\
0 & 0 & \sqrt{\frac{j-1}{j}}  & \frac{1}{\sqrt{j}} \\
0 & 0 & \frac{-1}{\sqrt{j}} & \sqrt{\frac{j-1}{j}} \\
\end{bmatrix}$$

The remaining steps apply the inverse operations to return the memory to its initial state and prepare it to receive the next pattern. The algorithm 
runs until all the patterns have been processed and stored on $\ket{M}$.

\subsubsection{The Retrieval Algorithm} \label{sc:ret}

The retrieval algorithm computes the Hamming distance between the input and all
the patterns superposed in the memory quantum state. It probabilistically
indicates the chance of a given input pattern being in the memory based on the
results of its distance distribution to the stored patterns in superposition.
If the input pattern is very distant from the patterns stored in the memory,
one will obtain $\ket{1}$ as a result with a significant probability. Otherwise, $\ket{0}$ would
be obtained. Since the memory state is in a superposition, the
retrieval algorithm can calculate the distances from input to all the patterns at once.

The PQM retrieval algorithm is described in Algorithm~\ref{alg:recover}.
It uses three quantum registers: $|i\rangle$, $|m\rangle$ and $|c\rangle$.
The pattern size gives the size of the first two registers and $|c\rangle$ is a single qubit register.
Step~\ref{rtv:load} of the algorithm loads the input pattern into the first register.
The second register $|m\rangle$ is the memory that contains all the stored patterns 
and  $|c\rangle$ is a control qubit initialized with a uniform superposition of $|0\rangle$ and $|1\rangle$.
The quantum state after the first step of the algorithm can be seen in Eq.~\eqref{eq:firstStateRetrieval}, where $\npat$ is the number of stored patterns.

\begin{algorithm}
    Load the input $\ip$ pattern in the quantum register $|i\rangle$ \\ \label{rtv:load}
    $\left| \psi_1 \right\rangle = \prod_{j=1}^n X_{m_j} \XOR_{i_j, m_j} \left| \psi_0 \right\rangle$ \\ \label{rtv:step1}
    $\left| \psi_2 \right\rangle = \prod_{i=1}^n \left(CU^{-2}\right)_{c,m_i} \prod_{j=1}^n U_{m_j} \left| \psi_1 \right\rangle$ \\ \label{rtv:step2}
    $\left| \psi_3 \right\rangle = H_c\prod_{j=n}^1  \XOR_{i_j, m_j}X_{m_j} \left| \psi_2 \right\rangle$\\ \label{rtv:step3}
    Measure qubit $\left|c\right\rangle$\\ \label{rtv:measure_c}
    \If{c == 0}{
        Measure the memory to obtain the desired state.
    }
    \caption{Probabilistic quantum memory retrieval algorithm}
    \label{alg:recover}
\end{algorithm}

\begin{equation}
\begin{split}
\ket{\psi_0} = \frac{1}{\sqrt{2\npat}} \sum_{k = 1}^{\npat} \ket{i;p^k;0} + \\
\frac{1}{\sqrt{2\npat}} \sum_{k = 1}^{\npat} \ket{i;p^k;1}
\end{split}
\label{eq:firstStateRetrieval}
\end{equation}

Step~\ref{rtv:step1} sets the $j$th qubit in memory register to $|1\rangle$, if the $j$th bit of input and memory are equal; and set to $|0\rangle$, if they differ. 
If a pattern stored in the memory is identical to the input pattern, step~\ref{rtv:step1} will set its memory state to  $|1\rangle_n$, where $n$ is the pattern size. 
In step~\ref{rtv:step2}, the operators $U$ (described in Eq.~\eqref{eq:U}) and controlled $U^{-2}$ are applied to the memory registers.

\begin{equation}
U = \left[\begin{array}{cc}
e^{(i\frac{\pi}{2n})} &   0 \\
0                  &   1
\end{array}\right]
\label{eq:U}
\end{equation}

Step~\ref{rtv:step2} is responsible for computing the Hamming distance between the input pattern $\ip$ and the patterns in the memory.
It computes the number of $0$s in the memory register (the qubits that differ between memory and input).
When $|c\rangle$ is $|0\rangle$, step~\ref{rtv:step2} calculates the number of $0$s in the memory state with a positive sign and with a negative sign when $|c\rangle$ is $|1\rangle$.

Step~\ref{rtv:step3} reverts the memory register to its original state by computing the inverse of step~\ref{rtv:step1}.
Step~\ref{rtv:measure_c} measures the quantum register $\ket{c}$.
An input pattern similar to the stored patterns increases the probability of measuring $|c\rangle = 0$, and an input that is very distant to the stored patterns increases the probability of measuring  $|c\rangle = 1$.
The measurement probabilities can be seen in Eq.~\eqref{eq:probs}, where $\npat$ is the number of stored patterns and $d_H(\ip,p^k)$ denotes the Hamming distance between input and the $k$th stored pattern.

\begin{equation}
\begin{split}
P(\ket{c} =\ket{0}) = \sum_{k = 1}^{\npat} \frac{1}{\npat} cos^2 \left(\frac{\pi}{2n}  d_H(\ip,p^k)\right)  \\
P(\ket{c} =\ket{1}) = \sum_{k = 1}^{\npat} \frac{1}{\npat} sin^2 \left(\frac{\pi}{2n}  d_H(\ip,p^k)\right)
\end{split}
\label{eq:probs}
\end{equation}

\section{The Quantum Weightless Classifier}\label{sc:QWC}

The Quantum Weightless Neural Network Classifier~\cite{silva2010weightless} (QWC) is composed of Probabilistic Quantum Memories acting as the network neurons.
The model is devised by using an array of PQM instances capable of distance-based classification.
Each PQM instance, by itself, works as a single class classifier, being responsible for the classification of just one of the classes in the dataset.
The model does not demand any training in the sense that the neurons do not have to be iteratively adjusted to learn from the training patterns.
The model classification algorithm and the necessary setup are detailed below.

\subsection{The Setup Algorithm}

The Quantum Weightless Classifier requires an initial setup algorithm in order to perform classification tasks.
For a given dataset with $n$ classes, the model has $n$ PQMs acting as neurons.
The training samples must be divided and grouped by class.
For each group, a new PQM is created and used to store all the samples belonging to that group, making in total $n$ PQM instances, one for each class.

The setup algorithm consists in storing the training samples on their respective PQM.
The $n$ PQMs together define a single classifier.
Algorithm~\ref{alg:pqmc_setup} describes the setup process.
Once all the training samples are correctly stored, the model can perform the classification task by calling the PQM retrieval algorithm.

\begin{algorithm}
    Initialize a PQM Classifier \label{line:1}\\
    \For{\textbf{each} class in dataset}{
        Create a new PQM and assign the class label to it \\
        Store the class training samples on the PQM \\
        Add the PQM to the PQM Classifier \\
        \label{line:4}}
    Return the PQM Classifier\label{line:7}
    \caption{Probabilistic Quantum Memory Classifier Setup}
    \label{alg:pqmc_setup}
\end{algorithm}

\subsection{The Classification Algorithm}

After the initialization described in Algorithm~\ref{alg:pqmc_setup}, the next module is
the classification algorithm shown in Algorithm~\ref{alg:pqmc_classification}.
In order to classify a new sample, the Quantum Weightless Classifier must 
present it to all the PQM neurons which constitute its network.
Each PQM neuron performs its retrieval algorithm using the presented sample as input.
Since each PQM holds the patterns of a specific class, each output will be the probability of the sample having similar features to the patterns of that specific class. Let $X$ be a random value corresponding to the number of $1$s in 
the probabilistic quantum memory output, the PQM neuron which outputs the 
smallest expected value, $E(X)$, is assumed to be the one that correctly 
classifies the sample.

\begin{algorithm}
    \For{\textbf{each} PQM in PQM Classifier}{
        Run the PQM retrieval algorithm with input $testPattern$ \\
        Calculate the expected value $E(X)$ from the retrieval algorithm output
        \\
    }
    Return the label from the PQM Classifier with the smallest
    $E(X)$\label{line:5}
    \caption{Probabilistic Quantum Memory Classifier Classification}
    \label{alg:pqmc_classification}
\end{algorithm}

\subsection{QWC Numeric Example}\label{sec:nex}

In this section, we present a numerical example of the PQM classifier evaluation. 
We show one limitation of the QWC with an artificial data set.
Assume the memory configuration in Table~\ref{Tab:PPQMex},  
and that the algorithm evaluates the input pattern $\ket{i}=\ket{0111010101}$. 
Patterns of memory $M_j$ in Table~\ref{Tab:PPQMex} have Hamming distance $j$ to the input pattern $\ket{i}$.  

\begin{table}
    \centering
    \begin{tabular}{|c|c|}\hline
         Memory 1 (M1)                                 & Memory 3 (M3) \\ \hline
         $m_{11}= \ket{0110010101}$      & $m_{31}= \ket{1111010110}$ \\
         $m_{12}= \ket{0101010101}$      & $m_{32}= \ket{1100010101}$ \\
         $m_{13}= \ket{0111010001}$      & $m_{33}= \ket{1101010001}$  \\
         $m_{14}= \ket{0011010101}$      & $m_{34}= \ket{1111100101}$  \\
        \hline
    \end{tabular}
    \caption{Memory configurations}\label{Tab:PPQMex}
\end{table}

The dataset has two classes $M_1$ and $M_3$ and 10-bit patterns. 
We suppose that $\ket{i} \in M_1$. 
To evaluate the classifier, we need to run the PQM retrieval algorithm for memories $M_1$ and $M_3$ with input $\ket{i}$ to obtain the probability of input pattern $\ket{i}$ to be in each memory.
So, we begin the algorithm obtaining the initial state in each memory:

\begin{align*}
\ket{\psi^{M_{1}}_0}=\frac{1}{\sqrt{4}}\Big(&\ket{0111010101;0110010101;C}\\
+ &\ket{0111010101;0101010101;C}\\
 +&\ket{0111010101;0111010001;C}\\+&
 \ket{0111010101;0011010101;C}\Big)
\end{align*}

\begin{align*}
\ket{\psi^{M_{3}}_0}=\frac{1}{\sqrt{4}}\Big(&\ket{0111010101;1111010101;C}\\+ &\ket{0111010101;1100010101;C}\\
+&\ket{0111010101;1101010001;C}\\+&\ket{0111010101;1111100101;C}\Big)
\end{align*}

Applying $\ket{\psi_1}$=$\prod_{j=1}^{n}\NOT_{m_j}\XOR_{i_{j}m_{j}}\ket{\psi_0}$, we obtain:

\begin{align*}
\ket{\psi^{M_{1}}_1}=\frac{1}{\sqrt{4}}\Big(&\ket{0111010101;1110111111;C}\\+ &\ket{0111010101;1101111111;C}\\
+&\ket{0111010101;1111111011;C}\\
+&\ket{0111010101;1011111111;C}\Big)
\end{align*}

\begin{align*}
\ket{\psi^{M_{3}}_1}=\frac{1}{\sqrt{4}}\Big(&\ket{0111010101;0111111100;C}\\+ &\ket{0111010101;0100111111;C}\\
 +&\ket{0111010101;0101111011;C}\\+&\ket{0111010101;0111001111;C}\Big)
\end{align*}

If we denote $d_{M_\alpha}(i,m_{\alpha k})$ as the Hamming distance between the input pattern $\ket{i}$ and every stored pattern $m_{\alpha k}$, then $d_{M_1}(i,m_{1k})=1$ and  $d_{M_3}(i,m_{3k})=3$,   $\forall k=1\dots 4$. After step~\ref{rtv:step2} of Algorithm~\ref{alg:recover} we obtain the state.

\begin{align*}
\ket{\psi^{M_{1}}_2}=\frac{1}{\sqrt{4}\sqrt{2}}\Big(&e^{(\frac{i\pi}{2n})}\ket{i;1110111111;0}\\+ &e^{(\frac{i\pi}{2n})}\ket{i;1101111111;0}\\ +&e^{(\frac{i\pi}{2n})}\ket{i;1111111011;0}\\+&e^{(\frac{i\pi}{2n})}\ket{i;1011111111;0}\Big)\\
+\frac{1}{\sqrt{4}\sqrt{2}}\Big(&e^{(\frac{-i\pi}{2n})}\ket{i;1110111111;1}\\+ &e^{(\frac{-i\pi}{2n})}\ket{i;1101111111;1}\\
 +&e^{(\frac{-i\pi}{2n})}\ket{i;1111111011;1}\\+&e^{(\frac{-i\pi}{2n})}\ket{i;1011111111;1}\Big).
\end{align*}

\begin{align*}
\ket{\psi^{M_{3}}_2}=\frac{1}{\sqrt{4}\sqrt{2}}\Big(&e^{(\frac{3i\pi}{2n})}\ket{i;0111111100;0}\\+ &e^{(\frac{3i\pi}{2n})}\ket{i;0100111111;0}\\
 +&e^{(\frac{3i\pi}{2n})}\ket{i;0101111011;0}\\+&e^{(\frac{3i\pi}{2n})}\ket{i;0111001111;0}\Big)\\
 +\frac{1}{\sqrt{4}\sqrt{2}}\Big(&e^{(\frac{-3i\pi}{2n})}\ket{i;0111111100;1}\\+& e^{(\frac{-3i\pi}{2n})}\ket{i;0100111111;1}\\
 +&e^{(\frac{-3i\pi}{2n})}\ket{i;0101111011;1}\\+&e^{(\frac{-3i\pi}{2n})}\ket{i;0111001111;1}\Big)
\end{align*}

Applying the step~\ref{rtv:step3} of Algorithm~\ref{alg:recover}, we obtain:

\begin{align*}
\ket{\psi^{M_1}_3}=\frac{1}{2}\Big[&\cos{\big(\frac{\pi}{20}\big)}\ket{i;m_{11};0}+\cos{\big(\frac{\pi}{20}\big)}\ket{i;m_{12};0}\\+&\cos{\big(\frac{\pi}{20}\big)}\ket{i;m_{13};0}+\cos{\big(\frac{\pi}{20}\big)}\ket{i;m_{14};0}\Big]\\
+\frac{1}{2}\Big[&\sin{\big(\frac{\pi}{20}\big)}\ket{i;m_{11};0}+\sin{\big(\frac{\pi}{20}\big)}\ket{i;m_{12};0}\\+&\sin{\big(\frac{\pi}{20}\big)}\ket{i;m_{13};0}+\sin{\big(\frac{\pi}{20}\big)}\ket{i;m_{14};0}\Big]
\end{align*}

\begin{align*}
\ket{\psi^{M_3}_3}=\frac{1}{2}\Big[&\cos{\big(\frac{3\pi}{20}\big)}\ket{i;m_{31};0}+\cos{\big(\frac{3\pi}{20}\big)}\ket{i;m_{32};0}\\+&\cos{\big(3\frac{\pi}{20}\big)}\ket{i;m_{33};0}+\cos{\big(\frac{3\pi}{20}\big)}\ket{i;m_{34};0}\Big]\\
+\frac{1}{2}\Big[&\sin{\big(\frac{3\pi}{20}\big)}\ket{i;m_{31};0}+\sin{\big(\frac{3\pi}{20}\big)}\ket{i;m_{32};0}\\+&\sin{\big(\frac{3\pi}{20}\big)}\ket{i;m_{33};0}+\sin{\big(\frac{3\pi}{20}\big)}\ket{i;m_{34};0}\Big]
\end{align*}

The next step measures the control qubit $\ket{C}$. 
Therefore, we must verify the probability amplitudes of $\ket{0}$ or $\ket{1}$. 
If only one memory $M_j$ returns 0, the pattern is classified as a member of the class $M_j$. 

$P^{M_1}(\ket{0})= cos^2{\left(\frac{\pi}{20}\right)} \approx 0.9755$ and 
$P^{M_3}(\ket{0})=cos^2{\left(\frac{3\pi}{20}\right)} \approx 0.79389$. 
The recognition probability of the input pattern $\ket{i}$ being recognized as a pattern of memory $M_1$ or $M_3$ is high for both memories, and then the model can perform a wrong classification.

A PQM classifier~\cite{silva2010weightless,schuld2014quantum} is not a good classifier in this situation. If we increase the size of $\ket{i}$ and $\ket{p^k}$ and if the input pattern  $\ket{i}$ have a Hamming distance of $1$ and $3$ for all stored patterns in $M_x$ 
and $M_y$, 
 respectively, the probability $P^{M_x}(\ket{0})$ 
 and $P^{M_y}(\ket{0})$ 
 will tend to $1$ and the algorithm would indicate that $\ket{i}$ is in both classes.
In the next section, we propose a parametric PQM that allows adjusting the model to a dataset.

\section{Parametric Quantum Weightless Classifier}\label{sec:pqwc}

The PQM retrieval algorithm output is a function of the Hamming distances between the input and the stored patterns.
This distance is not reliable for all datasets, as we have seen in the previous section. When patterns are 
close to patterns from another class, there is a high probability of misclassification.
In order to improve the memory output probabilities, we propose a modified version of the PQM.

We show how,	 for 
all memory size $ n \in \mathbb{N}$, 
that
it is possible to add a parameter $t$, with $0 < t \leq 1$, 
to the calculation of the probability in such a way that the distance between $P^{M_x}(\ket{0})$ and $P^{M_y}(\ket{0})$ 
will be sufficient to  assure a greater probability to perform a correct classification.

The idea is that given $d_x(i,m_x)<d_x(i,m_y)$, where $d_x$ and $d_y$ are Hamming distances between all patterns stored in
$M_x$ and $M_y$, respectively, we can add a parameter $t$ in the calculation of the probabilities and obtain values for 
$t\in (0,1)$ that increase the distance between the probabilities $P^{M_x}(\ket{0})$ and $P^{M_y}(\ket{0})$.

We want to make the probability 
$$P^{M_{x}}(\ket{0})= \sum_{k=1}^{p} \frac{1}{p}\cos^2{\Bigg(\frac{d_x\pi}{2nt}\Bigg)}$$
considerably close to 1, while the probability
$$P^{M_{y}}(\ket{0})= \sum_{k=1}^{p} \frac{1}{p}\cos^2{\Bigg(\frac{d_y\pi}{2nt}\Bigg)}$$
stays considerably close to zero. 

A Parametric Probabilistic Quantum Memory (P-PQM) operates as the PQM, but with the addition of a scale parameter in the retrieval algorithm. We replace the $U$ operator used in the PQM with the $U'$ operator described below, where $t$ is a free parameter. Note that for $t=1$ the P-PQM is equivalent to the PQM.

\begin{equation}
U' = \left[\begin{array}{cc}
e^{(i\frac{\pi}{2nt})} &   0 \\
0                  &   1
\end{array}\right]
\end{equation}

We define the Parametric Quantum Weightless Classifier (P-QWC), as the QWC classifier with a parametric retrieval algorithm.
In the next section, we revisit the example of the previous section to verify the effectiveness of the parametric approach.

\subsection{Numeric Example revisited} \label{sc:ex}

Taking the example in Section~\ref{sec:nex} as into account, we can apply the parameter from $\ket{\psi_2}$, hence:

\begin{align*}
\ket{\psi^{M_{1}}_2}=\frac{1}{\sqrt{4}\sqrt{2}}\Big(&e^{(\frac{i\pi}{2nt})}\ket{i;1110111111;0}\\+ &e^{(\frac{i\pi}{2nt})}\ket{i;1101111111;0}\\ +&e^{(\frac{i\pi}{2nt})}\ket{i;1111111011;0}\\+&e^{(\frac{i\pi}{2nt})}\ket{i;1011111111;0}\Big)\\
+\frac{1}{\sqrt{4}\sqrt{2}}\Big(&e^{(\frac{-i\pi}{2nt})}\ket{i;1110111111;1}\\+ &e^{(\frac{-i\pi}{2nt})}\ket{i;1101111111;1}\\
 +&e^{(\frac{-i\pi}{2nt})}\ket{i;1111111011;1}\\+&e^{(\frac{-i\pi}{2nt})}\ket{i;1011111111;1}\Big).
\end{align*}

\begin{align*}
\ket{\psi^{M_{3}}_2}=\frac{1}{\sqrt{4}\sqrt{2}}\Big(&e^{(\frac{3i\pi}{2nt})}\ket{i;0111111100;0}\\+ &e^{(\frac{3i\pi}{2nt})}\ket{i;0100111111;0}\\
 +&e^{(\frac{3i\pi}{2nt})}\ket{i;0101111011;0}\\+&e^{(\frac{3i\pi}{2nt})}\ket{i;0111001111;0}\Big)\\
 +\frac{1}{\sqrt{4}\sqrt{2}}\Big(&e^{(\frac{-3i\pi}{2nt})}\ket{i;0111111100;1}\\+& e^{(\frac{-3i\pi}{2nt})}\ket{i;0100111111;1}\\
 +&e^{(\frac{-3i\pi}{2nt})}\ket{i;0101111011;1}\\+&e^{(\frac{-3i\pi}{2nt})}\ket{i;0111001111;1}\Big)
\end{align*}

Applying step~\ref{rtv:step3} of Algorithm~\ref{alg:recover}, where $n=10$, we have:

\begin{align*}
\ket{\psi^{M_1}_3}=\frac{1}{2}\Big[&\cos{\big(\frac{\pi}{20t}\big)}\ket{i;m_{11};0}+\cos{\big(\frac{\pi}{20t}\big)}\ket{i;m_{12};0}\\+&\cos{\big(\frac{\pi}{20t}\big)}\ket{i;m_{13};0}+\cos{\big(\frac{\pi}{20t}\big)}\ket{i;m_{14};0}\Big]\\
+\frac{1}{2}\Big[&\sin{\big(\frac{\pi}{20t}\big)}\ket{i;m_{11};0}+\sin{\big(\frac{\pi}{20t}\big)}\ket{i;m_{12};0}\\+&\sin{\big(\frac{\pi}{20t}\big)}\ket{i;m_{13};0}+\sin{\big(\frac{\pi}{20t}\big)}\ket{i;m_{14};0}\Big]
\end{align*}

\begin{align*}
\ket{\psi^{M_3}_3}=\frac{1}{2}\Big[&\cos{\big(\frac{3\pi}{20t}\big)}\ket{i;m_{31};0}+\cos{\big(\frac{3\pi}{20t}\big)}\ket{i;m_{32};0}\\+&\cos{\big(3\frac{\pi}{20t}\big)}\ket{i;m_{33};0}+\cos{\big(\frac{3\pi}{20t}\big)}\ket{i;m_{34};0}\Big]\\
+\frac{1}{2}\Big[&\sin{\big(\frac{3\pi}{20t}\big)}\ket{i;m_{31};0}+\sin{\big(\frac{3\pi}{20t}\big)}\ket{i;m_{32};0}\\+&\sin{\big(\frac{3\pi}{20t}\big)}\ket{i;m_{33};0}+\sin{\big(\frac{3\pi}{20t}\big)}\ket{i;m_{34};0}\Big]
\end{align*}

We look for a value of $t$ which improves the probability of $\ket{i}$ to be recognized in memory $M_1$ and not recognized  in $M_3$.

As $P^{M_{1}}(\ket{0})= \cos^2{\left(\frac{\pi}{20t}\right)}$
and
$P^{M_{3}}(\ket{0})=\cos^2{\left(\frac{3\pi}{20t}\right)}.$

Let $f(t)$ be the function described in Eq.~\eqref{eq:ft}, where the first term is the one with the shortest Hamming distance to the input.

\begin{equation}f(t)=\cos^2{\left(\frac{\pi}{20t}\right)} - \cos^2{\left(\frac{3\pi}{20t}\right)}.\label{eq:ft}\end{equation}

\begin{figure}[t]
\centering
\includegraphics[width=0.7\columnwidth]{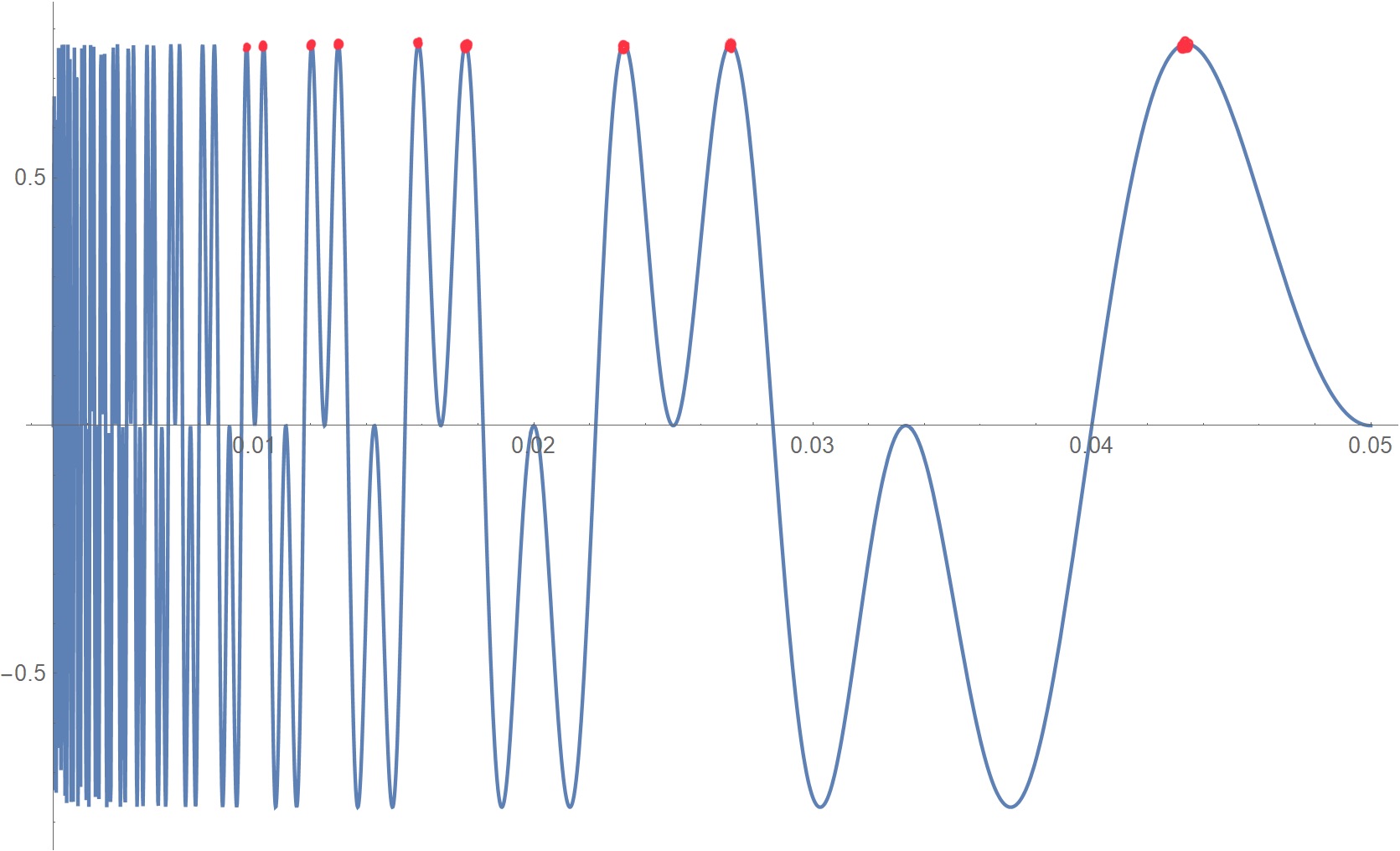}
\caption{Graph of $f(t)$ with maximum points.} 
\label{Rotulo1}
\end{figure}

Any value of $t$ that maximizes $f(t)$ is a parameter that makes the pattern recognition more reliable. 
It is sufficient to analyze the derivative of $f(t)$ to maximize $f$. By looking ath the maximum of $f$ we can get infinite values for $t$
that maximize the function, these values are those that give us the most significant distances between the probabilities. 
Below we will give some values of $t$ that maximizes $f(t)$. Values of $t$ where $f$ assumes minimum values lead to the classification of the input pattern $\ket{i}$ by the memory with the greatest Hamming distance. 
The graph of $f(t)$ is presented in Fig.~\ref{Rotulo1}.

We can obtain these values of $t$ by calculating the derivative of $f$. For our numeric example, we obtain 
$t\approx  \frac{0,0785398}{-0.238829+3.14158n}$, 
$t\approx \frac{0,0785398}{-0.133197+3.14158n}$,
$t\approx \frac{0,0785398}{0.238829+3.14158n}$, or
$t\approx \frac{0,0785398}{0.133197+3.14158n}, \forall n \in \mathbb{Z}.$

With $n=1$, we get values $t\approx 0.044$, $t\approx 0.017$, and $t\approx 0.0268$. Now let us calculate the probabilities for each value of $t$.

\begin{itemize} 
    \item [i.] $P^{M_{1}}(\ket{0})= \cos^2{\left(\frac{\pi}{20\cdot 0.044}\right)}\approx 0.83.$
    \item [i.] $P^{M_{3}}(\ket{0})= \cos^2{\left(\frac{3\pi}{20 \cdot 0.044}\right)}\approx 0.079.$
\end{itemize}

\begin{itemize} 
    \item [ii.] $P^{M_{1}}(\ket{0})= \cos^2{\left(\frac{\pi}{20 \cdot 0.017}\right)}\approx 0.96.$
    \item [ii.] $P^{M_{3}}(\ket{0})= \cos^2{\left(\frac{3\pi}{20 \cdot 0.017}\right)}\approx 0.084.$
\end{itemize}
\begin{itemize} 
    \item [iii.] $P^{M_{1}}(\ket{0})= \cos^2{\left(\frac{\pi}{20 \cdot 0.0268}\right)}\approx 0.86.$
    \item [iii.] $P^{M_{3}}(\ket{0})= \cos^2{\left(\frac{3\pi}{20 \cdot 0.0268}\right)}\approx 0.024.$
\end{itemize}

Note that in all cases the distance between the memories is considerably large, giving us greater confidence that the input pattern $\ket{i}$ is recognized by the memory $M_1$ since the Hamming distance associated with it is the smallest.

 We have just shown  that it is possible to obtain a parameter to improve the capabilities of the PQM classifier. Even in the case where the probabilities are very close and with a high value of $n$, the P-QWC can solve the cases where the QWC may not be efficient. Therefore, the P-QWC classifier provides a significant improvement over the QWC. In the next section, we use benchmark datasets to compare the QWC with the proposed P-QWC.

\subsection{Experiments} \label{sc:EVA}

We present in this section the experiments conducted on a conventional computer with a reduced classical version of Algorithm~\ref{alg:recover}.
To simulate the Probabilistic Quantum Memory classically we followed the description of its output probability described in Eq.~\eqref{eq:probs}.
Once the PQM classical representation is obtained, the QWC can be evaluated by following the setup and the classification algorithms described in Section~\ref{sc:QWC}. The same algorithm applies to the P-QWC model, with the only modification being the addition of a parameter to the Hamming distance calculation step.

To perform the experiments, we used categorical and numerical datasets from the UCI Machine Learning Repository~\cite{Dua:2017}.
Table~\ref{Tab:AgeWeight} presents the description of the selected datasets.
All datasets were preprocessed to binarize feature values and deal with any missing values.
Datasets containing real numerical values were not considered.
We replaced missing values by the value with the highest number of occurrences for the corresponding feature.

\begin{table*}[t]
    \centering
    \begin{tabular}{|c|c|c|c|c |}
        \hline
        Dataset & Classes  & Instances & Attributes & Missing Values \\
        \hline
        Balance scale & 3 & 625 & 4 & No \\
        Breast cancer&  2 & 286 & 9 & Yes \\
        Lymphography& 4 & 148 & 18 & No \\
        Mushroom & 2 & 8124 & 22 & Yes\\
        SPECT Heart & 2 & 267 & 22 & No  \\
        Tic-tac-toe &  2 & 958 & 9 & No \\
        Voting records& 2 & 435 & 16 & Yes  \\
        Zoo& 7 & 101 & 17 & No  \\
        \hline
    \end{tabular}
    \caption{Datasets characteristics}\label{Tab:AgeWeight}
\end{table*}

Following Algorithm~\ref{alg:pqmc_setup}, we stored the training samples in specific
PQMs according to the classes they belong to. Then, we follow Algorithm~\ref{alg:pqmc_classification}. The model classification accuracy was evaluated by
passing the patterns in the test set as input to each of the PQMs. The class
of the PQM which outputs the lowest expected value was set as the evaluated
pattern class. This algorithm was repeatedly applied to each of the evaluated datasets.
For the P-QWC, model we optimized and selected the parameters which achieved the best test set accuracy for each P-PQM in the classifier.
We tested 15 parameter values in the range $(0, 1]$ for each P-PQM.
In a quantum computer, it would be necessary to perform several runs to estimate the best value of $t$. In this simulated experiment, we performed only one run and analyzed the output state to determine $t$.

\subsection{Results}

\begin{figure}[t]
    \caption{Parameter impact on SPECT Heart dataset  \label{Fig:fig1}}
    \centering
    \includegraphics[width=0.8\textwidth]{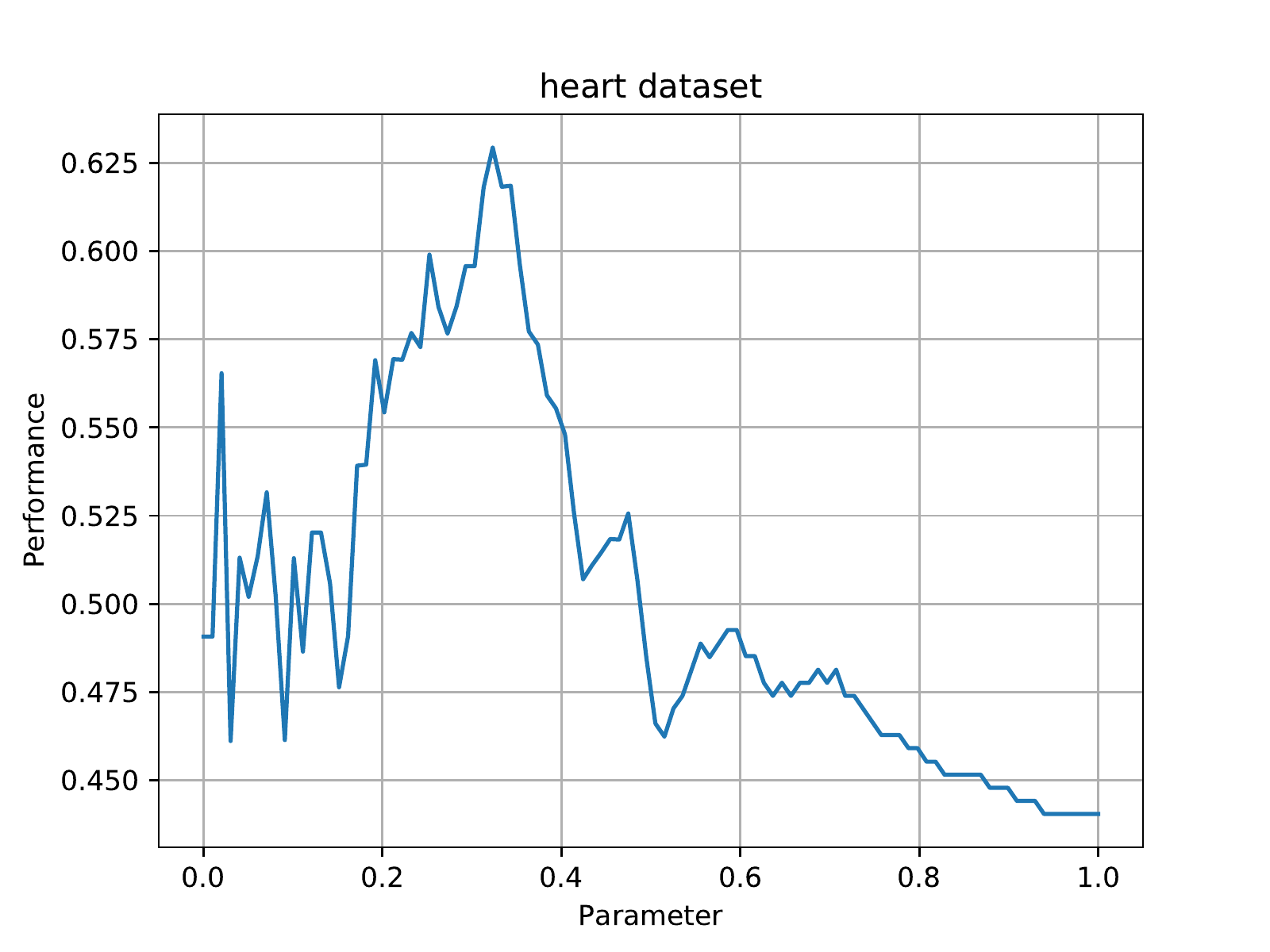}
\end{figure}

We verified that the P-QWC model performed better than the QWC over all the datasets.
The parameter influence on the SPECT Heart dataset performance can be seen in Fig.~\ref{Fig:fig1}.
The curve shows the performance obtained by using the same parameter value for all P-PQMs constituting the classifier.
Considering the original PQM performance is equivalent to P-PQM with parameter $1.0$, a considerable increase in classification performance
was observed through parameter variation and even better accuracies are possible by choosing different parameters for each P-PQM.

Table~\ref{Tab:results} describes the results obtained with the experimental setup described in Section~\ref{sc:EVA}.
The accuracy of the QWC and P-QWC models can be compared
against the results obtained using the k-nearest neighbors algorithm (KNN).
The accuracy values shown are the average obtained from 10-fold cross-validation. 
Standard deviations are included between parentheses. 
We choose KNN as a baseline comparison because it is also based on a measurement of distance. 
The KNN model was set to use uniform weights for all its
points and the k nearest neighbors value was optimized, and selected from values between 1 and 50.

A nonparametric statistical test was employed to perform an appropriate comparison of the models.
We used the Wilcoxon paired signed-rank test~\cite{demvsar2006statistical} with $\alpha=0.05$ to verify whether there exist significant differences between the compared classifiers performances over the chosen datasets. 
Statistically significant results between the P-QWC and KNN models are marked in bold, significant results in the comparison between the QWC and P-QWC models are italicized.
KNN and P-QWC are statistically equivalent in Balance scale, Breast cancer, SPECT Heart, Tic-tac-toe and Zoo datasets.
KNN has better accuracy on Mushroom and Voting records datasets.
P-QWC performed better on Lymphography dataset and also surpassed the QWC model in the SPECT Heart, Tic-tac-toe and Zoo datasets.
In comparison with QWC, P-QWC has better accuracy in all datasets with statistically significant results in datasets SPECT Heart, Tic-tac-toe and Zoo.

The P-QWC has a performance equivalent to KNN in five out of the eight tested datasets and outperforms it in one dataset.
The main advantage of the P-QWC is its memory requirements and the ability to receive inputs in superposition.
While a RAM node memory grows exponentially with input size~\cite{aleksander2009brief}, the QWC memory size grows linearly.
This memory advantage enables us  to implement new kinds of weightless neural networks architectures~\cite{silva2010weightless}.

\begin{table*}[ht]
	\centering
	\begin{tabular}{|c|c|c|c|}
		\hline
		Dataset & QWC & P-QWC  & KNN \\
		\hline
		Balance scale &  0.8111 (0.0666) & 0.87 (0.2512) &  0.8834 (0.0488)\\
		Breast cancer & 0.7309 (0.2639) & 0.7380 (0.2604) & 0.6970 (0.3797) \\
		Lymphography &  0.7829 (0.0815) & \textbf{0.8442} (0.1118) &  0.7695 (0.0931) \\
		Mushroom & 0.886 (0.0919) & 0.929 (0.073) & \textbf{1.0} (0.0) \\
		SPECT Heart & 0.4405 (0.2694) & \textit{0.8157} (0.1113) & 0.7921 (0.181) \\
		Tic-tac-toe & 0.4542 (0.1199) & \textit{0.8309} (0.0678) & 0.6714 (0.2989) \\
		Voting records & 0.892 (0.0575) & 0.8966 (0.0545) & \textbf{0.9332} (0.0377) \\
		Zoo & 0.92 (0.0872) & \textit{0.98} (0.06) & 0.96 (0.0663) \\
		\hline
	\end{tabular}
	\caption{10-fold cross-validation average accuracy per
		dataset}\label{Tab:results}
\end{table*}

\section{PQM for NISQ computers}
\label{sec:qubit-redc}

In this section, we first discuss why implementing the memory as proposed 
in~\cite{trugenberger2001probabilistic} would not be feasible on noisy small-scale quantum computers. 
Then, we propose a hybrid classical/quantum protocol implementation, optimized for devices with a reduced number of qubits. 
We performed an experiment on a small-scale quantum computer as a proof of concept that 
the modified retrieval algorithm will work on NISQ computers.

\subsection{Quantum-only implementation viability analysis}

The PQM retrieval algorithm calculates the distance of all the stored patterns to the input and outputs $|0\rangle$ with probability proportional to the given input pattern being close to patterns in the memory.
For patterns with $n$ bits, three quantum registers are needed for the quantum circuit: 
an input quantum register $|i\rangle$, a memory quantum register $|m\rangle$, both with $n$ qubits; 
and a controlled quantum register with at least one qubit. 
The retrieval algorithm can be described in 5 steps. 
There are $2n$ $CNOT$ and $NOT$ operations; $n$ $U$ and controlled $U^{-2}$ operations; one Hadamard operator;  and no operations involving more than two qubits. 
Operators $U$ and $CU^{-2}$ are not included in the set of gates available on the quantum device but can be constructed as a three gate composition. 
Thus, the significant issues with a quantum-only implementation are the memory scalability and the number of operations needed. 
A quantum memory used to store $n$-bit patterns will require $2n+2$ qubits, for the storage algorithm; and $2n+1$, for the retrieval algorithm. 
As the number of qubits is a limited resource on such small devices, the qubits requirement becomes a prohibiting issue. 
Furthermore, the high number of quantum operators results in too much noise being added to the output. 
These two issues make a quantum-only implementation on NISQ computers impractical. 
Therefore, due to the discussed limitations, we devised quantum-classical storage and 
retrieving algorithms to implement the PQM.

\subsection{Hybrid quantum-classical implementation method}

In the probabilistic quantum memory retrieval algorithm, the quantum register input always remains in a classical state. 
This fact and the hybrid classical/quantum architecture used in the actual quantum computers allows the removal of the input quantum register from Algorithm~\ref{alg:recover}. 

The main disadvantage of this approach is the need to recompile the circuit when the system receives an input, but as we show 
in~\cite{dosSantos2018quantum} the PQM has applications with a fixed input. 
Here, we remove the input quantum register and keep it in a classical variable. 
All the control operators from the input register to memory register are removed from the circuit and replaced with an $X$ operation applied to the 
$j$th memory bit only when the $j$th input bit is $1$. We also remove the $X$
gates from steps~\ref{rtv:step1} and~\ref{rtv:step3}, as they would cancel with the $X$ gates used to obtain 
$U = X Cu_1 X$. With these modifications, we can use the 5-qubit quantum computer to simulate a probabilistic quantum memory up to 4-bit patterns. 

In this work, we used the Quantum Information Software Kit
(QISKit) SDK~\cite{gadi_aleksandrowicz_2019_2562111} and run the circuit on the IBM Q Experience “Tenerife” 5-qubit 
quantum computer~\cite{ibmqx4spec}.  

The memory retrieval algorithm operations are further simplified by breaking down complex quantum operations into classical-quantum equivalent algorithms. 
These algorithms are conditioned on the classical input and apply a smaller number of gates to achieve the same resulting state. 

\subsection{Experiments}
\label{sec:exp}

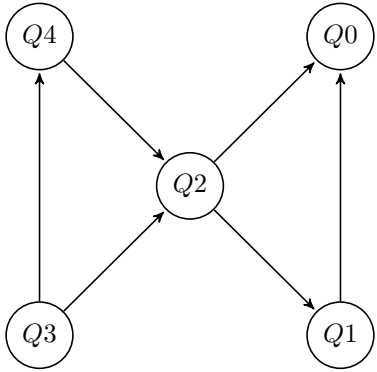
\begin{figure}[t]
	\center
	\begin{tikzpicture}[->,>=stealth',shorten >=1pt,auto,node distance=2.8cm,
	semithick]
	\tikzstyle{every state}=[fill=white, draw, text=black]
	
	\node[state] (A)                    {$Q2$};
	\node[state]         (B) [above right of=A] {$Q0$};
	\node[state]         (C) [below right of=A] {$Q1$};
	\node[state]         (D) [below left of=A] {$Q3$};
	\node[state]         (E) [above left of=A]  {$Q4$};
	
	\path 	(A) edge node {} (B)
	edge node {} (C)
	
	(C) edge node {} (B)
	
	(D) edge node {} (A)
	edge node {} (E)
	
	(E) edge node {} (A);
	\end{tikzpicture}
	\caption{IBM Q Experience ``Tenerife'' 5-qubit quantum computer topology}
	\label{fig:ibmqx4}
\end{figure}

The Tenerife 5-qubit quantum computer architecture does not have arbitrary qubits connections. 
All the possible connections in the Tenerife computer are defined by its architecture topology, 
which can be seen in Fig.~\ref{fig:ibmqx4}.

For this experiment, the PQM is tested with 2, 3 and 4 qubits memory size. We store different patterns on the PQM and run the retrieval algorithm using inputs with the same size as the memory patterns size. 
The PQMs are constructed by storing the control bit $c$ on the quantum register labelled as $Q3$ on Tenerife's topology. 
In this way, the necessary gates could be directly applied without swapping qubits. 
Table~\ref{tab:results} describes the results after 8192 executions of the 
probabilistic quantum memory running on the Tenerife backend and the local QISKit simulator. 
The calculated expected values for each memory state are included as well.

\begin{table*}[ht]
	\center
	\caption{Probabilistic quantum memory retrieval algorithm for input 
		$\ket{0}_{size}$ executed on the Tenerife backend with 8192 executions. The 
		results for 1, 2, 3, and 4-qubit PQM can be seen for different memory 
		configurations. The table shows the real output obtained from the 
		quantum 
		device, the results from the local simulator, and the expected 
		output 
		probability calculated numerically. We denote by $Ps$ the Pattern Size, and by $P_{T}, P_{Ls}$ and $P_{Ep}$ the probabilities obtained from the Tenerife backend, the local QISKit simulator and the calculated expected output, respectively.}
	\label{tab:results}
	\begin{tabular}{|l|l|c|c|c|}\hline
		Ps & Memory  state     &  
		\shortstack{$P_{T}(\ket{c}=\ket{0})$ } 		&
		\shortstack{$P_{Ls}(\ket{c}=\ket{0})$ } 		&
		\shortstack{$P_{Ep}(\ket{c}=\ket{0})$ } \\ \hline
		1 & 
		$\ket{0}$  																
		&
		0.9374 	& 1.0000	& 1.0000 \\ \hline
		1 & $\frac{1}{\sqrt{2}}(\ket{0} + \ket{1})$ 			& 0.5095 	& 
		0.4974 	& 0.5000  \\ \hline
		1 & 
		$\ket{1}$  																
		&
		0.0913 	& 0.0000	& 0.0000	\\ \hline
		2 & 
		$\ket{00}$    															
		&
		0.9033 	& 1.0000 	& 1.0000	\\ \hline
		2 & $\frac{1}{\sqrt{2}}(\ket{00} + \ket{01})$ 		& 0.6854 	& 
		0.7438 	& 0.7500	\\ \hline
		2 & 
		$\ket{11}$    															
		&
		0.1224 	& 0.0000 	& 0.0000	\\ \hline
		3 & 
		$\ket{000}$ 															
		&
		0.7618 	& 1.0000 	& 1.0000 \\ \hline
		3 & $\frac{1}{\sqrt{2}}(\ket{000} + \ket{010})$ 	& 0.6758 	& 
		0.8782	& 0.8750 \\ \hline
		3 & $\frac{1}{\sqrt{2}}(\ket{000} + \ket{100})$ 	& 0.6924 	& 
		0.8735 	& 0.8750 \\ \hline
		3 & $\frac{1}{\sqrt{2}}(\ket{000} + \ket{001})$ 	& 0.6246 	& 
		0.8804 	& 0.8750 \\ \hline
		3 & $\frac{1}{\sqrt{2}}(\ket{110} + \ket{111})$ 	& 0.2925 	& 
		0.1257 	& 0.125 \\ \hline
		3 & $\ket{111}$           												
		& 0.2278 	& 0.0000 	& 0.0000 \\ \hline
		4 & 
		$\ket{0000}$  															
		&
		0.7545 	& 1.0000 	& 1. 0000 \\ \hline
		4 & $\frac{1}{\sqrt{2}}(\ket{0000} + \ket{0100})$ & 0.7432 	& 0.9271	
		& 0.9268 \\ \hline
		4 & 
		$\ket{1000}$  															
		&
		0.7303	& 0.856 	& 0.8535 \\ \hline
		4 & $\frac{1}{\sqrt{2}}(\ket{0100} + \ket{1100})$ & 0.6844 	& 0.67 		
		& 0.6768 \\ \hline
		4 & 
		$\ket{1010}$  															
		&
		0.5441	& 0.4961 	& 0.5\\ \hline
		4 & $\frac{1}{\sqrt{2}}(\ket{0110} + \ket{1110})$ & 0.4830 	& 0.3336 	
		& 0.3232\\ \hline
		4 & 
		$\ket{1110}$  															
		&
		0.3827	& 0.1396 	& 0.1465\\ \hline
		4 & $\frac{1}{\sqrt{2}}(\ket{0111}+\ket{1111})$ 	& 0.2423 	& 
		0.073 	& 0.0732 \\ \hline
		4 & 
		$\ket{1111}$  															
		&
		0.2203 	& 0.0000 	& 0.0000\\ \hline
	\end{tabular}
\end{table*}

We calculate the expected outputs, obtained through numerical evaluation; the real outputs, obtained from the Tenerife backend; and the simulation outputs from the QISKit simulator. 
We calculate the percentual Mean Squared Error (MSE) of the expected outputs and the outputs of the experiment on the Tenerife backend. 
In the first set of experiments, we use a PQM 
with memory size equal to one and present all possible binary inputs. 
With memory state $\ket{0}$, $\ket{1}$ and 
$\frac{1}{\sqrt{2}}(\ket{0}+\ket{1})$, 
we obtain, respectively, MSE equal to $0.0058$, $0.0059$ and $0.0006$. 
The resulting mean error was $0.0041$.

In the second set of experiments, we use a 2-qubit PQM, we use all possible 
2-bit 
strings as inputs. With memory states $\ket{00}$, 
$\frac{1}{\sqrt{2}}(\ket{00}+\ket{01})$ 
and $\ket{11}$, we obtain, respectively, MSE equal to $0.0072$, $0.0027$ and 
$0.0066$.
The resulting mean error was $0.0055$. 

In the third set of experiments, we use a 3-qubit PQM and all possible 3-bit 
strings as inputs. With memory states $\ket{000}$, 
$\frac{1}{\sqrt{2}}(\ket{000}+\ket{010})$,
$\frac{1}{\sqrt{2}}(\ket{000} + \ket{100})$, $\frac{1}{\sqrt{2}}(\ket{000} + 
\ket{001})$, 
$\frac{1}{\sqrt{2}}(\ket{110} + \ket{111})$ and $\ket{111}$, we obtain, 
respectively, 
MSE equal to $0.0257$,  $0.0187$, $0.0133$, $0.023$, $0.0218$, and $0.026$.
The resulting mean error was $0.0214$. 

\begin{figure}[t]
	\subfloat[Results for input pattern $0000$\label{subfig:4q_0}]{%
		\includegraphics[width=0.5\textwidth]{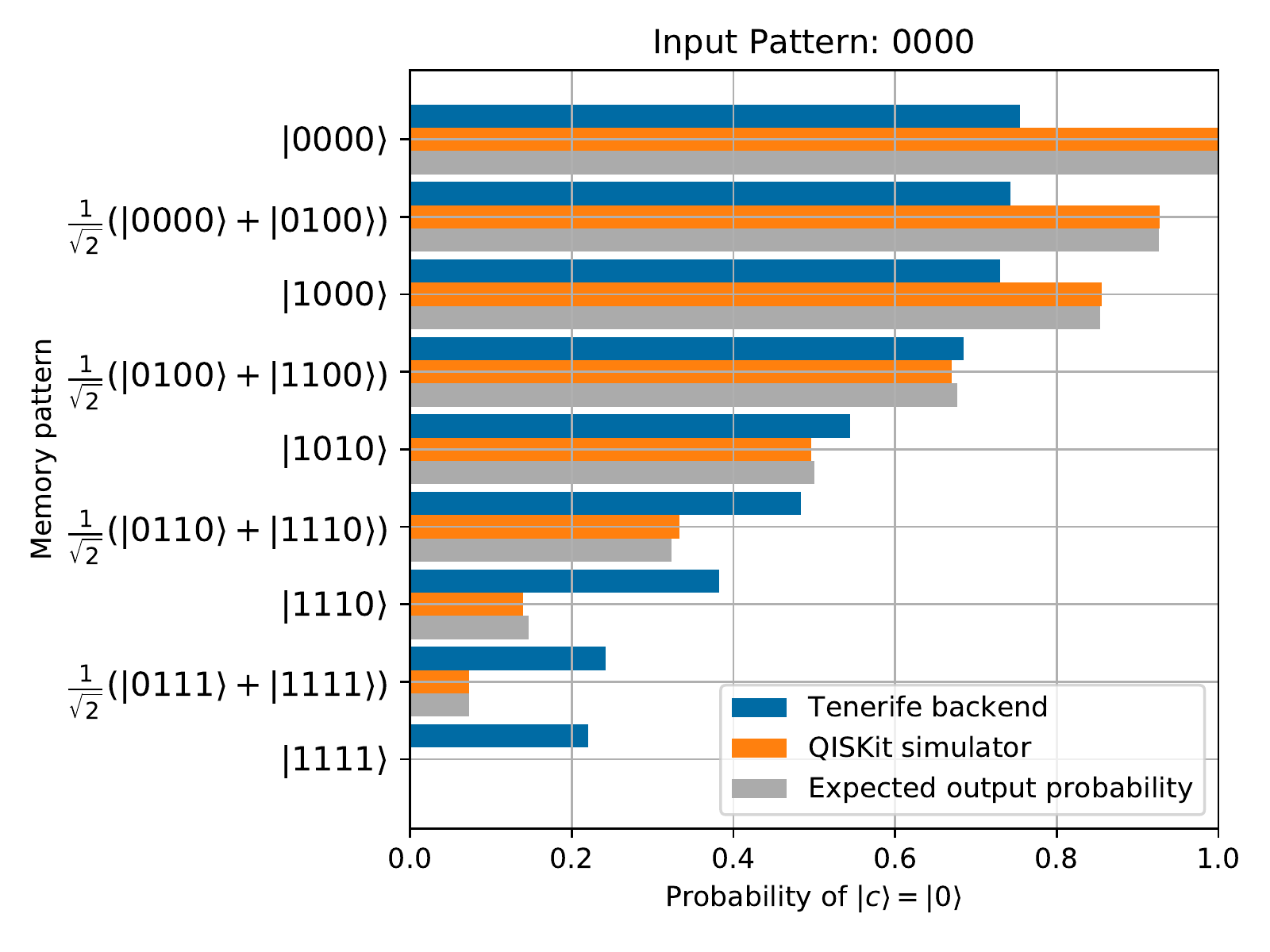}%
	}\hfil
	\subfloat[Results for input pattern $1111$\label{subfig:4q_15}]{%
		\includegraphics[width=0.5\textwidth]{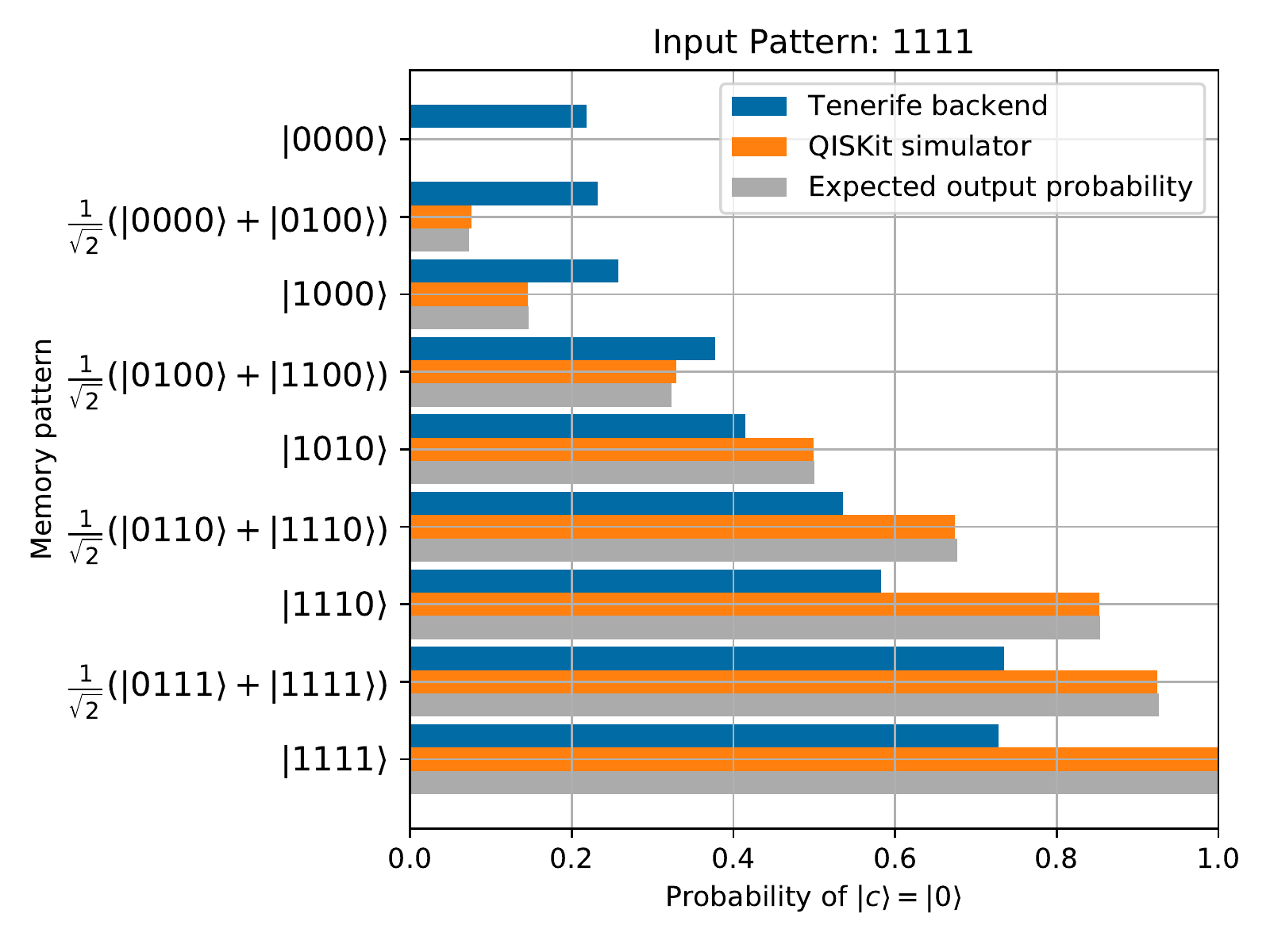}%
	}
	\caption{Results obtained from executing the retrieval algorithm of a 4-qubit PQM
		on the Tenerife backend and the QISKit local simulator, including the
		numerically calculated expected probabilities.}
	\label{fig:4q}
\end{figure}

Finally, in the fourth set of experiments we use a 4-qubit PQM and all possible 
4-bit 
strings as inputs. With memory states $\ket{0000}$, 
$\frac{1}{\sqrt{2}}(\ket{0000} + \ket{0100})$,
$\ket{1000}$, $\frac{1}{\sqrt{2}}(\ket{0100} + \ket{1100})$, $\ket{1010}$, 
$\frac{1}{\sqrt{2}}(\ket{0110} + \ket{1110})$, $\ket{1110}$, 
$\frac{1}{\sqrt{2}}(\ket{0111}+\ket{1111})$
and $\ket{1111}$, we obtain, respectively, 
MSE equal to $0.0228$, $0.0174$, $0.0236$, $0.0144$, $0.0228$, $0.0148$, 
$0.0248$, $0.0141$, and $0.0226$.
The resulting mean error was $0.0197$. Results for inputs 0000 and 1111 are 
displayed in Fig.~\ref{fig:4q}.

Although the calculated outputs do not correspond precisely to the predicted outputs, the probability of obtaining $|0\rangle$ from a measurement of the quantum register $\ket{c}$ is still related to the distance between the input and the memory content. 
In all experiments, the estimate of $\ket{c}$ can be used to verify if a pattern is close to the patterns in the memory.

The proposed simplified retrieval 
algorithm of the PQM was successfully implemented in the Tenerife architecture without swapping quantum bits.
We conjecture that a quantum computer with planar architecture where $n$ qubits 
are connected to a single qubit (to be used as a quantum register $\ket{c}$) can 
be used to efficiently implement an $n$-qubits PQM in near term quantum 
computers.

\section{Conclusion} \label{sc:conclusion}

In this work, we proposed a parametric version of the PQM named P-PQM.
We performed an empirical evaluation of a quantum weightless classifier and proposed a modification which achieved a considerable improvement in the classification capabilities of the model.
The proposed parametric quantum model used as a classifier (P-QWC) performed better or equivalent to its unmodified version (QWC) in all datasets used in the experiments.

We also presented a modification of the PQM to allow its implementation on Noisy Intermediate-Scale 
Quantum computers. As a proof of concept, the model was implemented in a small-scale quantum computer.
We verified through experiments that a noisy version of the PQM can be implemented on a 5-qubit quantum computer.

There are several possible future works. We can adapt the model to use different distance functions, use different architectures for the quantum weightless classifier, investigate how to reduce noise on the P-PQM, and define an error cost function to update the parameter $t$ as in a variational quantum circuit.

\section*{Code avaiability} All code used in this work can be made available upon reasonable request.

\section*{Acknowledgements}
    This work was supported by the Serrapilheira Institute (grant number Serra-1709-22626), CNPq Edital Universal (grants numbers \mbox{420319/2016-6} and {421849/2016-9} and FACEPE (grant number IBPG-1578-1.03/16). We acknowledge use of the IBM Q for this work. The views expressed are those of the authors and do not reflect the official policy or position of IBM or the IBM Q team.

\bibliography{mybibfile}

\end{document}